\documentclass[prb,reprint,amsmath,amssymb,showpacs,twocolumn]{revtex4}
\usepackage{dcolumn}
\usepackage{amsmath}
\usepackage{epsf}
\usepackage{graphicx}
\usepackage{bm}
\usepackage{color}

\begin{document} 

\preprint{APS/123-QED}
\title{Nonlocal correlations in a proximity-coupled normal metal}

\author{Taewan Noh$^1$}
\author{Sam Davis$^2$}
\author{Venkat Chandrasekhar$^{1,2}$}
\affiliation{$^1$Department of Physics and Astronomy, Northwestern University, Evanston, IL 60208, USA, \\ 
$^2$Graduate Program in Applied Physics, Northwestern University, Evanston, IL 60208, USA}
\date{\today}

\begin{abstract}
We report evidence of large, nonlocal correlations between two spatially separated normal metals in superconductor/normal-metal (SN) heterostructures, which manifest themselves as nonlocal voltage generated in response to a driving current.  Unlike prior experiments in SN heterostructures, the nonlocal correlations are mediated not by a superconductor, but by a proximity-coupled normal metal.  The nonlocal correlations extend over relatively long length scales in comparison to the superconducting case.  At very low temperatures, we find a reduction in the nonlocal voltage for small applied currents that cannot be explained by the quasiclassical theory of superconductivity, which we believe is a signature of new long-range quantum correlations in the system. 
\end{abstract}


\maketitle

\section{Introduction}

Electrons in spatially separated normal metals in mutual contact with a superconductor show correlations that are evidence of quantum entanglement due to their interaction with the Cooper pairs in the superconductor. \cite{byers, deutscher, falci, feinberg, melin, brinkman, morten, yeyati} Experimentally, the correlations manifest themselves as a nonlocal voltage that develops on one normal metal in response to a current injected from another normal metal into the superconductor.
In such a hybrid structure, three processes that contribute to the nonlocal voltage have been investigated. Two, crossed Andreev reflection (CAR) and elastic co-tunneling (EC), arise from the interaction of quasiparticles in the spatially separated normal metals that are mediated by the Cooper pairs in the superconductor, and decay spatially on the scale of the superconducting coherence length $\xi_S$, typically about 100 nm for Al.  \cite{feinberg, brinkman}  The third process, charge imbalance,\cite{clarke, tinkham} is associated with the conversion of a quasiparticle current into a supercurrent in the superconductor, and decays over the charge imbalance length $\Lambda_{Q^*}$, which is of the order of a few microns.\cite{arutyunov, brauer, hubler} 
None of these three nonlocal processes are expected in a normal metal with an induced superconducting proximity effect, where there is no superconducting order parameter.
Consequently, it comes as a surprise that large nonlocal signals can indeed be observed when a quasiparticle current is injected into a normal metal that is proximity coupled to two superconductors.  The experimental manifestation of these nonlocal signals is very similar to that observed in normal-metal/superconductor/normal-metal (NSN) structures,\cite{beckmann, russo, cz, cz1, cz2,arutyunov, brauer, hubler} although their origin appears to be quite different. 

\section{Experiment}
\begin{figure}[h!]
      \includegraphics[width=6cm]{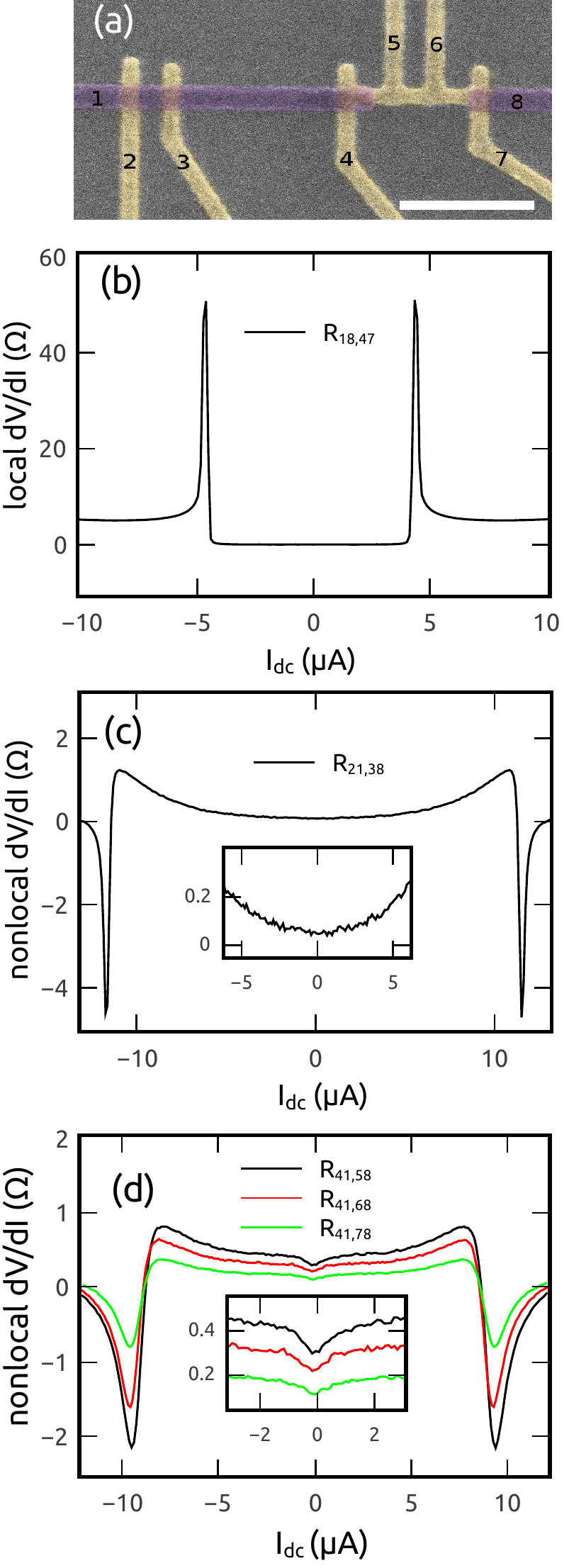}
      \caption{(a) False color scanning electron micrograph of the sample discussed in the text.  Light areas are Au; the darker lines are Al.  The numbers mark the contacts used for four-terminal differential resistance measurements. The size bar is
                500 nm. (b)  Local differential resistance as a function of the applied current $I_{dc}$.  (c) Nonlocal resistance of the NSN configuration of sample shown in (a) as a function of $I_{dc}$. 
                Inset: Expanded view of the zero bias region. (d) Nonlocal resistance of the proximity-coupled normal metal shown in (a) as a function of $I_{dc}$. Inset: Expanded view of the zero bias region. All the measurements shown here are performed at 24 mK.}
      \label{fig1}
     \vspace{-0.75cm}
\end{figure}

The samples measured in this work were fabricated by using conventional electron beam lithography. We selected Au as the normal metal and Al as the superconductor. O$_2$ plasma etching was performed prior to the deposition of Al
on top of Au to ensure transparent interfaces. Figure \ref{fig1}(a) shows an example which involves two different
measurement configurations. The left side of the sample corresponds to a NSN structure similar to that measured in previous experiments.\cite{cz, cz1} In the configuration on the right side of the sample, the superconductor between the two normal metals is replaced by a proximity-coupled normal metal.
The length $L$ of the normal metal wire between the two superconductors was designed so that the two superconductors were Josephson-coupled in the temperature range of interest.  
Given the multiprobe nature of the sample, we shall use the notation $R_{ij,kl}=dV_{kl}/dI_{ij}$ to denote the four-terminal differential resistance, where the ac and dc currents are applied between contacts $i$ and $j$, and the resulting ac voltage measured between contacts $k$ and $l$, 
with the contacts numbered as in Fig. \ref{fig1}(a). 
Figure \ref{fig1}(b) shows the local differential resistance $R_{18,47}$ and demonstrates that the two superconductors are indeed Josephson-coupled through the normal metal arm, with a critical current $I_c \sim 3.6$ $ \mu$A at $T=$ 24 mK.

Figure \ref{fig1}(c) shows a nonlocal measurement on the left part of the sample of Fig. \ref{fig1}(a) at 20 mK, corresponding to the NSN configuration measured previously,\cite{cz, cz1} and it can be seen that the resulting curves are also similar to what was observed earlier:  at $I_{dc}=0$, there is a small but finite resistance, of the order of a few milliohms.  This contribution arises from the difference between EC and and CAR.  EC is expected to give rise to a positive differential resistance, while CAR is expected to give rise to a negative contribution.  The positive sign of the zero bias resistance implies that the EC contribution is larger than the CAR contribution, as was seen in earlier experiments.\cite{cz,cz1}  At larger values of $|I_{dc}|$ ($\sim$ 10 $\mu$A), a peak in resistance is observed that is associated with charge imbalance.\cite{cz,cz1} This is followed by a sharp dip down to negative values in the differential resistance.  Similar negative resistance dips are seen in a variety of 
NS and ferromagnet-superconductor (FS) structures at dc currents near the critical current of the superconductor.\cite{cz1}  The origin of this negative differential resistance is not entirely clear:  the fact that they are seen at values of $I_{dc}$ close to the critical current, and they do not scale with length as expected from EC/CAR \cite{cz1} suggest that they are associated with nonequilibrium effects rather than nonlocal EC or CAR, a point of view also supported by theoretical calculations.\cite{bergeret}    

Figure \ref{fig1} (d) shows the corresponding nonlocal resistance measurements on the configuration on the right side of the sample of Fig. \ref{fig1}(a); the difference again is that in this configuration, the superconductor that forms the bridge between the two normal metal parts in the conventional NSN configuration is now replaced by a proximity-coupled normal metal.  The overall shape of the curves is very similar to the curve in Fig. \ref{fig1}(c), with a finite positive differential resistance at $I_{dc}=0$, a peak in differential resistance at  $|I_{dc}|\sim 8$ $\mu$A, followed by a drop in differential resistance to negative values at higher $|I_{dc}|$.  As was found for conventional NSN devices,\cite{cz} the nonlocal differential resistance also decreases as the length of the $V+$ voltage lead from the current path increases.  However, there are some significant differences between the data of Fig. \ref{fig1}(d) and Fig. \ref{fig1}(c).  First, the magnitude of the zero bias differential resistance 
is larger in Fig. \ref{fig1}(a) in comparison to the NSN configuration of Fig. \ref{fig1}(c), and also in comparison to previous work on NSN devices\cite{cz}.  Second, and more significant, there is small dip in the differential resistance near $I_{dc}=0$ in the nonlocal measurements that is absent in the conventional NSN measurements.  The inset to Fig. \ref{fig1}(d) shows that this dip also scales with the distance of the $V+$ nonlocal voltage probe from the current path.  The dip is evidence of new nonlocal correlations that exist in this system.    

\begin{figure}
      \begin{center}
      \includegraphics[width=7cm]{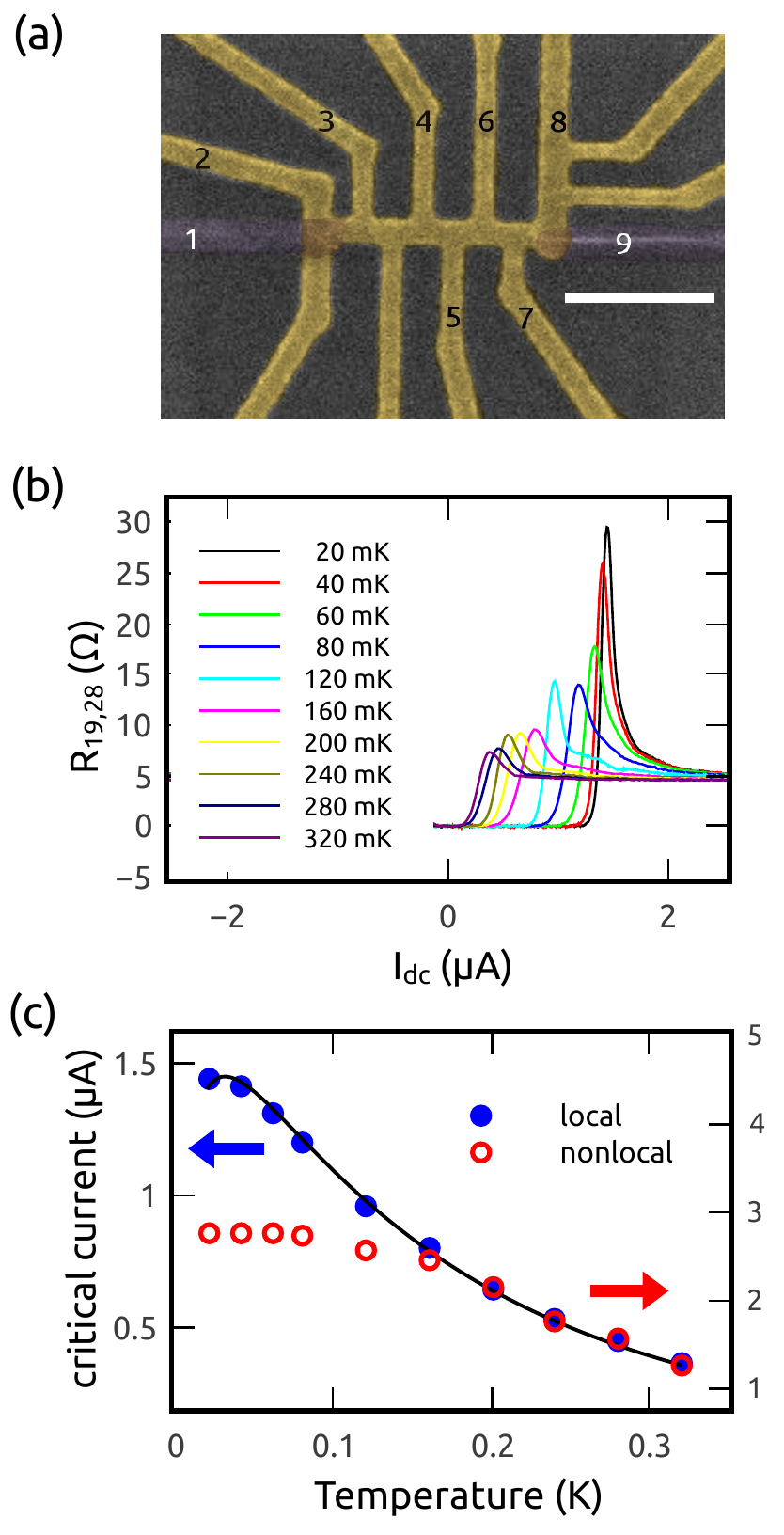}
      \caption{(a) False color scanning electron micrograph of the sample discussed in the text.  Light areas are Au; the darker lines are Al.  The numbers mark the contacts used for four-terminal differential resistance measurements. The size bar is
                500 nm. (b)  Local differential resistance $R_{19,28}$ as a function of the applied current $I_{dc}$.  (c)  Blue circles show the measured critical current, determined by the position of the peak in the differential resistance in (b), as a function of temperature $T$.  The solid line shows a fit to the expected temperature dependence for a long SNS junction of length $L$.\cite{dubos}   Open red circles are the finite values of $I_{dc}$ at which the nonlocal differential resistance $R_{31,49}$ (Fig. \ref{fig3}(a)) has its minimum.  } 
      \label{fig2}
     \end{center}
     \vspace{-0.75cm}
\end{figure}

In order to perform a more detailed examination of the length dependence of the nonlocal differential resistance in this new sample configuration, we fabricated devices with multiple terminals.  An image of one of these devices is shown in Fig. \ref{fig2}(a).  While a number of samples were measured and showed similar results, for the remainder of the paper, we shall concentrate on this sample, for which we have the most complete data.   

For this sample, the electronic diffusion coefficient $D=(1/3) v_F \ell$ = 110 cm$^2$/s, as determined from resistance measurements of the normal metal wires above the critical temperature $T_c \sim 1.15$ K of the Al (here $v_F$ is the Fermi velocity and $\ell$ is the elastic scattering length in the gold).  The distance $L$= 0.75 $\mu$m between the two NS interfaces in Fig. \ref{fig2}(a) gives a Thouless energy $E_c = \hbar D/L^2$  (the relevant energy scale for the superconducting proximity effect) of 11.7 $\mu$eV, with a corresponding Thouless length $L_T =\sqrt{\hbar D/k_B T}$ = 293 nm/$\sqrt{T}$.   This corresponds well with the fact that a finite supercurrent was observed below $T \sim$ 0.7 K.  

Figure \ref{fig2}(b) shows a measurement of the local differential resistance $R_{19,28}$ as a function of $I_{dc}$ at various temperatures, corresponding to the differential resistance of the SNS junction.  The position of the peaks in this curve identify the critical current $I_c$.  The blue circles in Fig. \ref{fig2}(c) show the $T$ dependence of $I_c$, and the solid line is a fit to the functional form, $I_c = B T^{3/2} \exp (- A/L_T)$ (where A and B are constants), the form expected for a SNS junction in the long junction limit \cite{dubos} ($\Delta>>E_c$, where $\Delta$ is the gap in the superconductor).  Although the fit is quite good, $I_{c}$ at base temperature is smaller than the value $I_{c0}=10.82 E_c/eR_N$ predicted for a simple SNS junction\cite{dubos} ($R_N$ is the normal state resistance).  For this sample, $I_{c0} e R_N / E_c \sim 0.56$, with $R_N$ = 4.56 $\Omega$.  In multiterminal structures,  $I_c$ is expected to be suppressed due to a modification of the induced minigap.\cite{virtanen}  
If we assume that the measured $I_{c0}$ is related to an effective Thouless 
energy $E_c^*$ by the 
same 
theoretical prediction 
for a simple SNS 
junction, we obtain  $E_c^*$ =(0.56/10.82)$\times$11.7 = 0.60 $\mu$eV.

\begin{figure}
      \begin{center}
      \includegraphics[width=6.5cm]{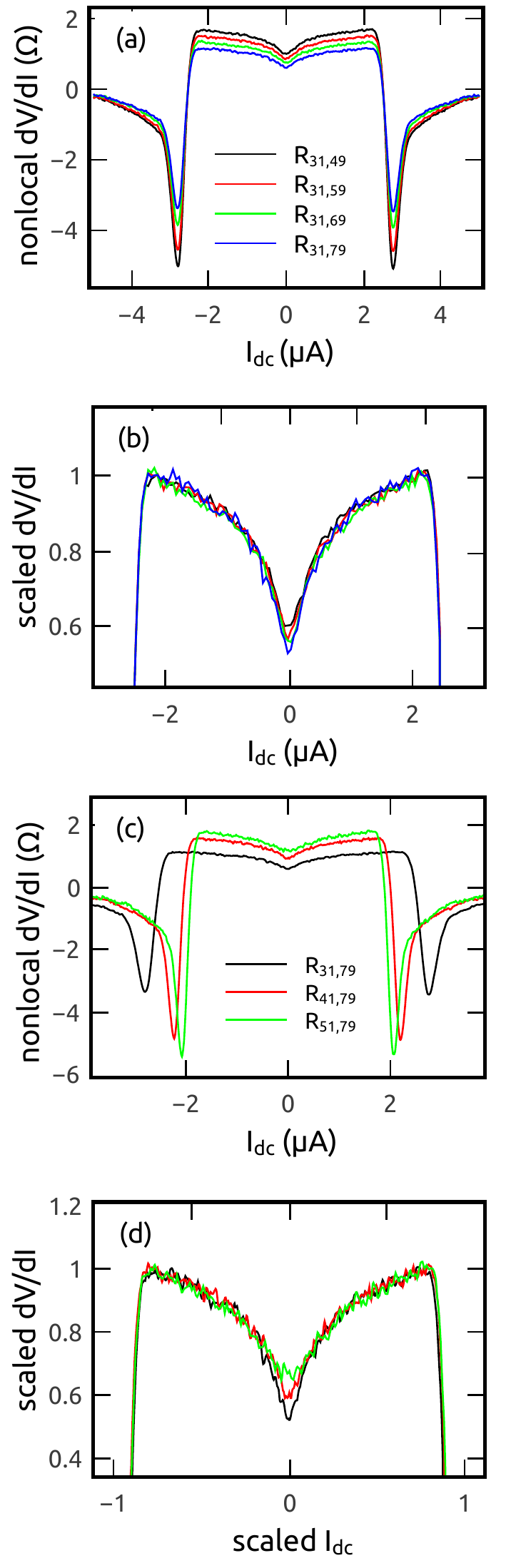}
      \caption{ (a) Nonlocal resistances as a function of dc current bias for 4 nonlocal  configurations, $R_{31,49}$, $R_{31,59}$, $R_{31,69}$ and $R_{31,79}$. (b)  Data of (a), with the curves for $R_{31,49}$, $R_{31,59}$, $R_{31,69}$ and $R_{31,79}$ scaled so that their normalized peaks at $\pm 2.3$ $\mu$A match.  (c)  Nonlocal resistances $R_{31,79}$, $R_{41,79}$ and $R_{51,79}$, where the current injection terminal 
is changed, but the voltage contacts remain the same. (d) Data of (c), with the curves for $R_{31,79}$, $R_{41,79}$, and $R_{51,79}$ with both $x$ and $y$ axes scaled as described in the text. 
}      
      \label{fig3}
     \end{center}
     \vspace{-0.75cm}
\end{figure}

We now discuss the nonlocal measurements.  Figure \ref{fig3}(a) shows the nonlocal $dV/dI$ at 20 mK for four different configurations, each with the current sourced through normal electrode 3 and drained through a superconducting electrode 1.  (For the electrode numbers, please refer to Fig. \ref{fig2}(a).)  The overall shape of the resulting traces is similar to what was observed in the first sample (Fig. \ref{fig1}(d)):  at $I_{dc}$ = 0, nonlocal $dV/dI$ is finite and grows with current, resulting in a peak at a finite current of $\sim 2.3$ $\mu$A, after which there is a sharp drop to negative values before it goes to zero at high bias.  The nonlocal $dV/dI$ also decreases as the distance of the $V+$ contact from the current path increases.  Finally, there is a sharp dip in $dV/dI$ near $I_{dc}$ = 0 that is not present in NSN samples.

In earlier nonlocal NSN experiments,\cite{cz,cz1} it was found that the zero bias differential resistance $R_{nl}(0)$ and the peak at finite $I_{dc}$ decayed with the distance from the current injection electrode, but with different length scales: While the zero bias resistance  decayed exponentially with $\xi_S$ as expected from CAR/EC, the peak was associated with charge imbalance and found to decay linearly with $\Lambda_{Q^*}$.  In the current experiments, $R_{nl}(0)$ and the peak resistance also scale differently with distance.  Figure \ref{fig3}(b) shows the  curves of Fig. \ref{fig3}(a) scaled along the $y$-axis so that their peaks at finite bias match.  With this scaling, the curves match over most of the range of current, except near zero bias.  (The inset to Fig. \ref{fig3}(b) shows an expanded version of the zero bias nonlocal differential resistance.)  This shows clearly that $R_{nl}(0)$ and the finite bias resistance scale differently with length, as was found for the NSN samples.
In the NSN case, $R_{nl}(0)$ decayed exponentially on a length scale of $\xi_S$: Here one might expect that $R_{nl}(0)$ should decay exponentially with $L_T$.  Contrary to expectation, both $R_{nl}(0)$ and the peak resistance scale linearly with the distance from the current injection point, although the slopes are different (data not shown), reflecting the different scaling evident in Fig. \ref{fig3}(b). 

Similar behavior is observed if one keeps the voltage contacts the same, but moves the contact used to inject the current.  Figure \ref{fig3}(c) shows these data.  As before, the magnitude of the nonlocal differential resistance increases with decreasing distance from the $V+$ probe to the current path.  Unlike the data in Fig. \ref{fig3}(a), however, the position of the negative resistance dips at finite dc bias also changes.  As we discuss later, injecting a dc current through a normal contact into the proximity-coupled normal metal wire induces a supercurrent between the two superconductors.  The negative differential resistance is associated with exceeding the critical current of the SNS junction: by changing the current path, the fraction of the injected current that is converted into supercurrent changes, and hence the position of the negative resistance dips in terms of the injected current also changes.  However, if we scale the $x$-axis so that the position of the dips match, and independently scale 
the $y$-axis so that the magnitude of the resistance peaks at finite bias match, we obtain the curves shown in Fig. \ref{fig3}(d).  Again, this demonstrates clearly that the resistance dip at zero bias scales differently with length in comparison to the finite bias part of the differential resistance.     

\section{Theoretical Analysis}

\begin{figure}
      \begin{center}
      \includegraphics[width=6.5cm]{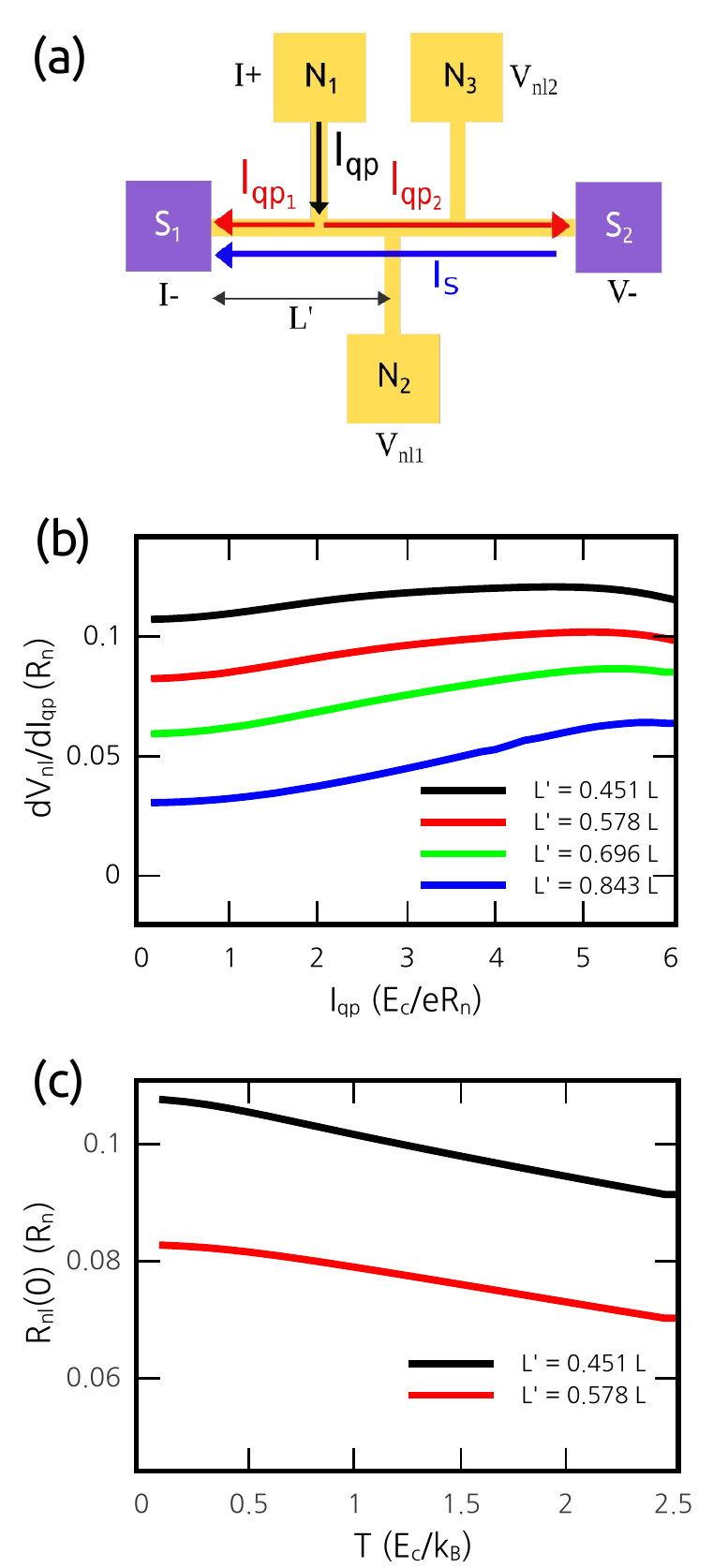}
      \caption{ (a)  Schematic of the current separation model.  The injected quasiparticle current $I_{qp}$ splits into two currents, $I_{qp1}$ and $I_{qp2}$, which go towards the two superconducting contacts $S_1$ and $S_2$ respectively.  $I_{qp2}$ is compensated by a counterflowing supercurrent $I_s$ that flows from $S_2$ to $S_1$.  (b)  Results of the numerical simulations based on the quasiclassical theory of superconductivity for the nonlocal resistances for the geometry of (a) as a function of $I_{qp}$. The distance $L'$ between the normal probe (on the normal metal) and $S_1$ is $0.451L$, $0.578L$, $0.696L$ and $0.843L$, where $L$ is the length of normal metal between $S_1$ and $S_2$. The position of $N_1$ is fixed at $0.196L$ from $S_1$.   These values are chosen to match the sample geometry.  (c)  Calculated zero bias resistance as a function of temperature.    
}      
      \label{fig4}
     \end{center}
     \vspace{-0.85cm}
\end{figure}

What is the origin of this behavior?  In SNS structures with Josephson coupling between two superconducting electrodes, supercurrents and quasiparticle currents can coexist in a proximity-coupled normal metal over distances much longer than $\xi_S$.\cite{baselmans, shaikhaidarov, crosser}  As the superconducting electrodes $S_1$ and $S_2$ are Josephson-coupled at low enough temperatures, they are at the same electrochemical potential, which we take to be 0 here for simplicity.  A finite potential $V$ applied to the current injection electrode $N_1$ will drive a quasiparticle current $I_{qp}
$ into the proximity-coupled normal metal as shown in Fig. \ref{fig4}(a).  Since $S_1$ and $S_2$ are both at zero potential, this quasiparticle current will split into two components:  $I_{qp1}$ will flow towards $S_1$, and $I_{qp2}$ will flow towards $S_2$, the ratio $I_{qp1}/I_{qp2}$ being determined by the inverse ratio of the resistances of the normal sections between the current injection point and $S_{1,2}$.  As $S_2$ is a voltage probe, no net current can flow into it.  Hence, $I_{qp2}$ must be balanced by a counterflowing supercurrent $I_S$ that flows from $S_2$ to $S_1$ such that $I_S= - I_{qp2}$ and $I_S + I_{qp1} = I_{qp}$, in turn giving rise to a phase difference $\phi$ between $S_1$ and $S_2$.  This situation will persist until $I_S$ exceeds $I_c$ of the junction.  Since $I_{qp1}$ is always less than $I_{qp}$, the injected dc current $I_{qp}$ at which this occurs is always greater than the critical current $I_c$.   Evidence for this model can be seen by examining the 
value of current at which this occurs.  The red open circles in Fig. \ref{fig2}(c) show the value of $I_{dc}$ as a function of $T$ at which the minimum in the resistance at finite bias is observed in the nonlocal measurement as shown in Fig. \ref{fig3}(a).  The ordinate axis has been scaled so that the data points lie on top of the measured $I_c$ (blue circles) at high temperatures.  At lower temperatures, however, they show a weaker temperature dependence.  This latter behavior has been observed previously by other groups,\cite{crosser} and arises from the reduction of $I_c$ due to the nonequilibrium quasiparticle distribution introduced by $I_{qp}$.\cite{baselmans}

$I_{qp2}$ will result in a finite ``nonlocal'' voltage between the normal voltage contacts $N_{2,3}$ and the second superconductor $S_2$ that will be proportional the resistance of the normal wire between $N_{2,3}$ and $S_2$, plus the resistance of the interface $NS_{2}$.  Thus, one would expect to see the linear scaling mentioned earlier.  In addition, as the current injection point is moved closer to $S_2$, one would expect to see an increase of the nonlocal voltage (and hence differential resistance) and decrease of $I_{dc}$ where minimum in the resistance appears as the ratio $I_{qp2}/I_{qp}$ increases.  Figure \ref{fig3}(c) which shows the nonlocal differential resistance for a fixed voltage lead but different current injection points, confirms this. 

Hence, it appears that the current separation model describes well the nonlocal resistance that we observe.  However,  closer analysis of the current bias and temperature dependence of the nonlocal resistance reveals some significant discrepancies.  Quantitative predictions for the differential resistance can be obtained by simultaneously solving numerically the Usadel equations and the kinetic equations\cite{venkat}  to obtain the nonlocal $dV_{nl}/dI_{qp}$ as a function of $I_{qp}$, and $R_{nl}(0)$ as a function of temperature. 
We used the public domain numerical solvers developed by Pauli Virtanen,\cite{pauli} based on the Riccati parametrization of the quasiclassical equations for superconductors in the diffusive limit.\cite{belzig,pauli2}  The starting point for the simulations is the equation for the total current
\begin{equation}
\mathbf{j}(\mathbf{R}, T) = e N_0 D \int dE  [M_{33} (\partial_{\mathbf{R}} h_T) + Q h_L + M_{03} (\partial_{\mathbf{R}} h_L)]. 
\label{eqn1}
\end{equation}
Here $N_0$ is the electronic density of states at the Fermi energy, $D$ is the diffusion coefficient, $h_T$ and $h_L$ are the tranverse and longitudinal quasiparticle distribution functions, which in equilibrium at energy $E$ in reservoirs at a potential $V$ have the form
\begin{equation}
h_{L,T} = \frac{1}{2} \left[ \tanh \left(\frac{E+eV}{2k_B T}\right) \pm  \left(\frac{E-eV}{2k_B T}\right) \right],
\label{eqn2}
\end{equation}
the spectral supercurrent $Q$ is given by

\begin{equation}
Q=\Re \left( 2 N^2 [\gamma \nabla \tilde{\gamma} - \tilde{\gamma} \nabla \gamma]\right),
\label{eqn3}
\end{equation}
and the dimensionless diffusion coefficients $M$ by
\begin{equation}
M_{33}=|N|^2 (|\gamma|^2 + 1)(|\tilde{\gamma}|^2 + 1)
\label{eqn4}
\end{equation}
and
\begin{equation}
M_{03}=|N|^2 (|\tilde{\gamma}|^2 - |\gamma|^2),
\label{eqn5}
\end{equation}
where $N=(1+\gamma \tilde{\gamma})^{-1}$.  $\gamma$, $\tilde{\gamma}$ are the Riccati parametrization parameters that satisfy the coupled equations
\begin{eqnarray}
&&D \nabla ^2 \gamma - 2 N \tilde{\gamma} |\nabla \gamma |^2 + 2 i E \gamma =0
\label{eqn6} \nonumber\\
&&D \nabla ^2 \tilde{\gamma} - 2 N \gamma |\nabla \tilde{\gamma} |^2 + 2 i E \tilde{\gamma} =0
\label{eqn7}
\end{eqnarray}
in the normal metal wires.  The boundary conditions for the differential equations are that $\gamma$ and $\tilde{\gamma}$ are zero at a normal reservoir, while on a superconducting reservoir
\begin{eqnarray}
\gamma^R &=& - \frac{\Delta}{E+ i\sqrt{|\Delta|^2 - (E+i\delta)^2}} \nonumber\\
\tilde{\gamma}^R &=&  \frac{\Delta^*}{E+ i\sqrt{|\Delta|^2 - (E+i\delta)^2}},
\label{eqn8}
\end{eqnarray}
where $\Delta = |\Delta| e^{i\phi}$ is the complex order parameter in the superconducting reservoir, $\phi$ being the phase of the superconductor.  

The first term on the RHS in Eq. (\ref{eqn1}) is the quasiparticle current, the second term is the supercurrent, and the third term corresponds to conversion of quasiparticle to supercurrent, and is typically negligible in a normal metal.  At the nodes, the Riccati parameters are continuous, and their derivatives sum to zero.  The spectral charge and energy currents also sum to zero along a node, and are conserved in a normal wire. 

To calculate the differential resistance as a function of the current, we use the sample geometry in Fig. \ref{fig4}(a).  This has two superconducting reservoirs and three normal reservoirs.  Using one normal reservoir for current injection, the other two normal reservoirs are modeled as voltage contacts, allowing us to calculate two nonlocal resistances simultaneously.  A voltage $V$ is applied to the normal contact $N_1$ with the superconducting reservoirs $S_1$ and $S_2$ being at zero potential, and the resulting current $I_{qp}$ flowing from $N_1$ is calculated under the boundary condition that no current flows into contacts $N_2$, $N_3$ and $S_2$.  In order to satisfy these boundary conditions, the voltages $V_{nl1}$ and $V_{nl2}$ on the normal contacts $N_2$ and $N_3$ and the phase difference $\phi$ between the two superconducting contacts $S_1$ and $S_2$ are adjusted in an iterative loop.  As $I_{qp}$ increases, a quasiparticle current $I_{qp2}$ flows into $S_2$, which is counterbalanced by a 
supercurrent $I_S$ that flows from $S_2$ to $S_1$.  As $I_{S}$ approaches the critical current, the simulations have greater difficulty converging, and we have shown in the figures only that range of $I_{qp}$ over which the boundary 
conditions are satisfied.  For the plot, the nonlocal differential resistances $dV_{nl1}/dI_{qp}$ and $dV_{nl2}/dI_{qp}$ are calculated numerically.  Taking the length of the wire between the superconducting reservoirs to be $L$, the nonlocal resistances shown in the simulations of Fig. \ref{fig4}(b) correspond to the positions of the normal leads in the actual sample shown in Fig. \ref{fig2}(a), with the length of each normal reservoir from the proximity coupled normal wire being 0.75 $L$.  We have also performed simulations with different values of this length, and this value most closely resembles the shape of the experimental curve.  The temperature of the simulations correspond to 20 mK. For the temperature dependence, a similar simulation was done, but only a small voltage $\pm V$ was applied to $N_1$, and the resulting values used to calculate the differential resistances as a function of temperature.  Ideally, we would have liked to have a sufficient number of normal contacts to model exactly the 
device of Fig. \ref{fig3}(a).  However, the numerical calculation does not converge fast enough over all values of dc current with a larger number of reservoirs.  In order to model the length dependence of the experimental data, we have therefore repeated the calculations with the normal reservoirs placed at different lengths along the normal wire, corresponding to the measured dimensions of the experimental sample. 

\begin{figure}
      \begin{center}
      \includegraphics[width=7.5cm]{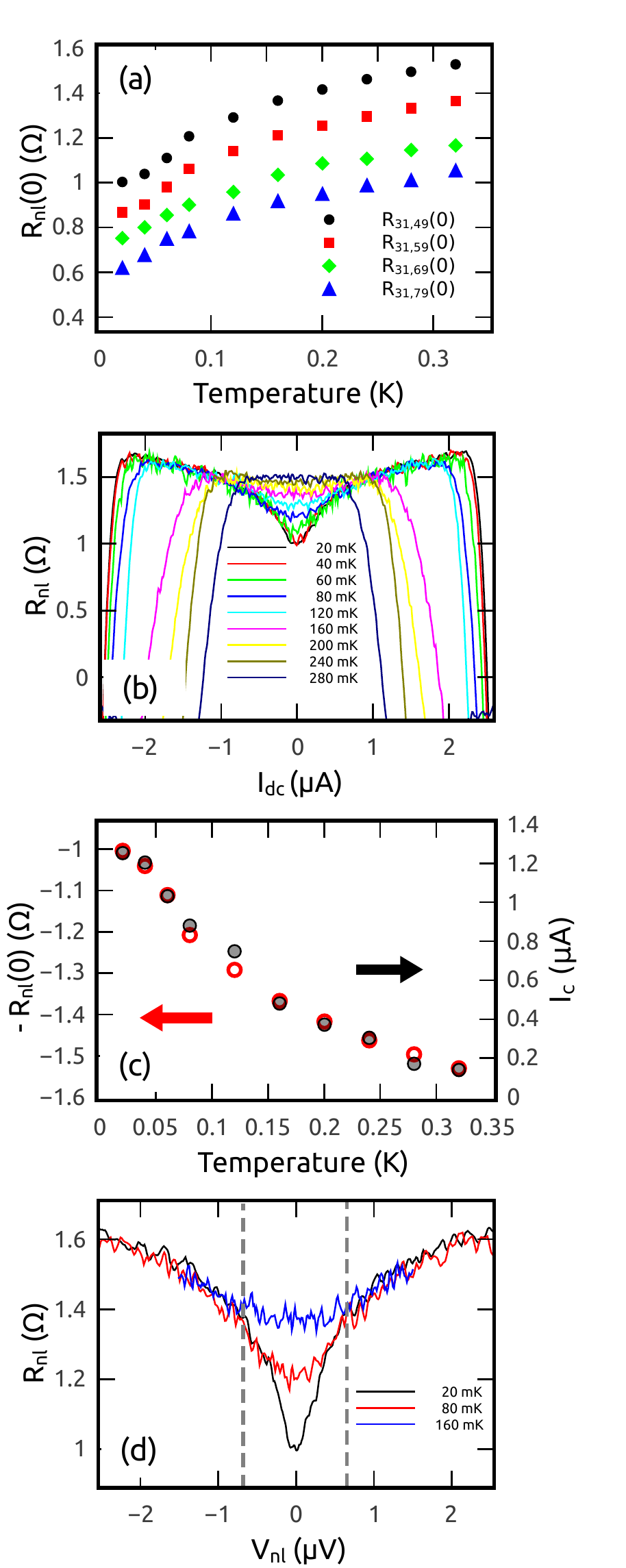}
      \caption{ (a)  Temperature dependence of $R_{nl}(0)$ for all four nonlocal resistance configurations.  (b)  Nonlocal differential resistance for the closest configuration at a number of different temperatures. (c) Temperature dependence of $- R_{nl}(0)$ for the configuration in (b) and temperature dependence of
       the local critical current $I_c$. 
      (d)  Data from (b) at 20 mK, 80 mK and 160 mK, plotted as a function of the nonlocal voltage $V_{nl}$ obtained by numerical integration. Dotted lines are guides to the eyes for the points where 20 mK and 80 mK data deviate significantly from 160 mK data.
}      
      \label{fig5}
     \end{center}
     \vspace{-0.75cm}
\end{figure}

The resulting curves are shown in Fig. \ref{fig4}(b) and \ref{fig4}(c) respectively. 
There are some significant differences between the results of the simulations and our measurements, but we would first like to emphasize that the results of the simulations are consistent with previous experimental and theoretical results,\cite{courtois,petrashov} under the assumption that the nonlocal differential resistance we measure is just the resistance of the appropriate length in Fig. \ref{fig4}(a) of the proximity-coupled normal metal.  Consider first Fig. \ref{fig4}(b), which shows that the nonlocal differential resistance increases as the injected quasiparticle current $I_{qp}$ is increased from zero.  As we noted above, injecting a finite $I_{qp}$ results in a finite supercurrent flowing between the two superconductors $S_1$ and $S_2$, corresponding to a finite phase difference $\phi$ between them.  At the values of $I_{qp}$ in Fig. \ref{fig4}(a), $\phi < \pi/2$.  The resistance of the proximity-coupled normal metal metal is a function of $\phi$, being a minimum at $\phi = 0$.  Hence, as $I_{qp}$ 
and thus $\phi$ increases, $R_{nl}$ would be expected to increase, as seen in Fig. \ref{fig4}(b).     

Figure \ref{fig4}(c) shows that the zero bias nonlocal differential resistance $R_{nl}(0)$ is expected to increase as $T$ decreases.  This behavior is simply the well-known reentrance effect of a proximity-coupled normal metal \cite{courtois}:  As the temperature is decreased below the transition temperature of the superconductor, the resistance of the proximity-coupled normal metal first decreases, but then reaches a minimum at a certain temperature $T_0$, increasing as the temperature is decreased further.  For a single N wire connected to a S reservoir, $T_0 \sim 5 E_c/k_B$.\cite{venkat}  For more complicated geometries, $T_0$ may be modified, and a self-consistent calculation as we have done is required.  These calculations show that experimentally we are in the regime $T<T_0$, with the resistance rising as the temperature is decreased, since $L_T \ge L$ in the temperature regime of interest.

We now compare the simulations to the experimental data.  First, note that the nonlocal differential resistance in the simulations of Fig. \ref{fig4}(b) appears to saturate as $I_{qp}\rightarrow 0$.  This is in contrast to the experimental data, which show a sharp dip in the differential resistance at zero bias.  Second, the temperature dependence of the zero bias nonlocal resistance is exactly opposite that that predicted by the simulations.  Figure \ref{fig5}(a) shows $R_{nl}(0)$ for the four nonlocal configurations of Fig. \ref{fig3}(a) in the low temperature regime.  The nonlocal resistance \textit{decreases} with decreasing temperature, in direct contrast to the behavior expected from Fig. \ref{fig4}(c).  At higher temperatures, the nonlocal resistance eventually goes down to zero, which is due to the vanishing of the supercurrent between the superconducting electrodes, without which no nonlocal signal can be observed.

A clue to the origin of the decrease in $R_{nl}(0)$ with decreasing $T$ can be seen in the temperature evolution of $dV/dI$ vs. $I_{dc}$, which is shown in Fig. \ref{fig5}(b) for one nonlocal configuration.  Apart from the decrease in $I_c$ with increasing $T$, the only major difference between the different temperature traces is the growth of the dip at zero bias.  In order to study the temperature dependence of this feature, we plot $-R_{nl}(0)$ together with the measured $I_c$ of the SNS junction in Fig. \ref{fig5}(c), with appropriate offset and scaling in the $y$-axis so that the data points coincide at higher temperatures.  $-R_{nl}(0)$ matches the exponential behavior of $I_c$ at higher temperatures, but the two curves diverge at lower temperatures, with $I_c$ saturating, but $-R_{nl}(0)$ still showing a strong exponential dependence.  Measurements of the critical current involve sending a substantial dc current through the sample, which may cause some heating at the lowest temperatures, depending on 
how the critical current is defined.  In the absence of this heating,  -$R_{nl}$ follows the temperature dependence of $I_c$, which is directly related to $E_c^*$.
 
Further evidence that the dip near zero bias is related to the effective Thouless energy $E_c^*$ can be found by integrating the nonlocal differential resistance to obtain $dV/dI$ vs $V_{nl}$.  Figure \ref{fig5}(d) shows the 20, 80 and 160 mK  from Fig. \ref{fig5}(b) plotted in this manner.  The voltage at which the 20 mK curve deviates significantly from the 80 and 160 mK curves at low bias-- the voltage at which the zero bias dip starts developing -- is approximately 0.6 $\mu$V,
much smaller than $E_c/e$, but in very good agreement with the effective Thouless energy $E_c^*/e$ defined earlier. 

\section{Conclusion}

We emphasize again that the dip in the nonlocal differential resistance that we observe at low temperatures is \emph{not} described by the conventional quasiclassical theory of superconductivity. 
For nonlocal measurements on conventional NSN devices near zero bias, CAR is expected to give a negative contribution to the nonlocal resistance,\cite{deutscher} as two electrons with energies less than $\Delta$, one from each normal metal, combine to form a Cooper pair.  The decrease in the nonlocal resistance in our samples suggests a similar process is happening in these samples, since the pair coupling in the proximity coupled  normal metal is finite.  Thus, two electrons, one from each normal lead, combine to form a correlated pair in the proximity coupled normal metal.  Exactly how this process occurs is not clear, as it is not described by our current understanding of nonequilibrium transport in proximity-coupled  normal metals.

In summary, measurements on proximity-coupled normal metals reveal a signature of long-range nonlocal quasiparticle correlations that may be related to the formation of pair correlations in the proximity-coupled normal metal.  Further study is required to elucidate the origin of these correlations.  

\section*{ACKNOWLEDGMENT}

We thank P. Virtanen for help with his quasiclassical superconductivity numerical code.  This research was supported by the NSF under grant No. DMR-1006445.


\begin{thebibliography}{text}
\bibitem{byers} J. M. Byers and M. E. Flatt\'e, {Phys. Rev. Lett.} \textbf{74}, 306 (1995).
\bibitem{deutscher} G. Deutscher and D. Feinberg, {Appl. Phys. Lett.} {\bf 76}, 487 (2000).
\bibitem{falci} G. Falci, D. Feinberg and F. W. J. Hekking,  {Europhys. Lett.} {\bf 54}, 255 (2001).
\bibitem{feinberg} D.Feinberg, {Euro. Phys. J. B} {\bf 36}, 419 (2003).
\bibitem{melin} R. M\'{e}lin and D. Feinberg, {Phys. Rev. B} {\bf 70}, 174509 (2004).
\bibitem{brinkman} A. Brinkman and A. A. Golubov, {Phys. Rev. B} {\bf 74}, 214512 (2006).
\bibitem{morten} J. P. Morten, A. Brataas and W. Belzig, {Applied Physics A} {\bf 89}, 609 (2007).
\bibitem{yeyati} A. L. Levy Yeyati, F. S. Bergeret,  A. Mart\'{i}n-Rodero and T. M. Klapwijk, {Nature Physics} {\bf 3}, 455 (2007).
\bibitem{clarke} J. Clarke, {Phys. Rev. Lett.} {\bf 28}, 1363 (1972). 
\bibitem{tinkham} M. Tinkham and J. Clarke, {Phys. Rev. Lett.} {\bf 28}, 1366 (1972).
\bibitem{brauer} J. Brauer, F. H\"{u}bler, M. Smetanin, D. Beckmann and H. v. L\"{o}hneysen, {Phys. Rev. B} {\bf 81}, 024515 (2010).
\bibitem{hubler} F. H\"{u}bler, J. Camirand Lemyre, D. Beckmann and H. v. L\"{o}hneysen, {Phys. Rev. B} {\bf 81}, 184524 (2010).
\bibitem{arutyunov} K. Yu. Arutyunov, H. -P. Auraneva and A. S. Vasenko, {Phys. Rev. B} {\bf 83}, 104509 (2011).
\bibitem{beckmann} D. Beckmann, H. B. Weber and H. v. L\"{o}hneysen, {Phys. Rev. Lett.} {\bf 93}, 197003 (2004).
\bibitem{russo} S. Russo, M. Kroug, T. M. Klapwijk and A. F. Morpurgo, {Phys. Rev. Lett.} {\bf 95}, 027002 (2005).
\bibitem{cz} P. Cadden-Zimansky and V. Chandrasekhar, {Phys. Rev. Lett.} {\bf 97}, 237003 (2006).
\bibitem{cz1} P. Cadden-Zimansky, Z. Jiang and V. Chandrasekhar, {New J. Phys} {\bf 9}, 116 (2007).
\bibitem{cz2} P. Cadden-Zimansky, J. Wei, V. Chandrasekhar, {Nat. Phys.} {\bf 5}, 393 (2009).
\bibitem{bergeret} F. S. Bergeret, A. Levy Yeyati, {Phys. Rev. B.} {\bf 80}, 174508 (2009).
\bibitem{dubos} P. Dubos, H. Courtois, B. Pannetier, F. K. Wilhelm, A. D. Zaikin and G. Sch\"{o}n,  {Phys. Rev. B} {\bf 63}, 064502 (2001).
\bibitem{virtanen} P. Virtanen, PhD Thesis, Helsinki University of Technology (2009).
\bibitem{baselmans} J. J. A. Baselmans, A. F. Morpurgo, B. J. van Wees and T. M. Klapwijk, {Nature} {\bf 397}, 43 (1999).
\bibitem{shaikhaidarov} R. Shaikhaidarov, A. F. Volkov, H. Takayanagi, V. T. Petrashov, and P. Delsing, {Phys. Rev. B} {\bf 62}, (R)14649 (2000).
\bibitem{crosser} M. S. Crosser, J. Huang, F. Pierre, P. Virtanen, T. T. Heikkil\"{a}, F. K. Wilhelm and N. O. Birge, {Phys. Rev. B} {\bf 77}, 014528 (2008).
\bibitem{lambart} C. J. Lambert and R. Raimondi, { J. Phys.: Condens. Matter} {\bf 10}, 901 (1998).
\bibitem{venkat} V. Chandrasekhar, chapter in {Physics of Superconductors}, vol II, eds. Bennemann and Ketterson, Springer-Verlag, (2004).
\bibitem{pauli}Virtanen, P.,  \url{http://ltl.tkk.fi/~theory/usadel1.html}.
\bibitem{belzig} W. Belzig, F. K. Wilhelm, C. Bruder, G. Sch\"{o}n and A. D. Zaikin, { Superlattices Microstruct} {\bf 25}, 1251 (1999).
\bibitem{pauli2}P. Virtanen, P. and T.T. Heikkil\"a,  \textit{Appl. Phys. A} {\bf 89}, 625 (2007).
\bibitem{courtois} H. Courtois, Ph. Gandit, D. Mailly and B.  Pannetier, {Phys. Rev. Lett.} {\bf 76}, 130 (1996).
\bibitem{petrashov}V.T. Petrashov, R. Sh. Shaikhaidarov, I.A. Sosnin, P. Delsing, T. Claeson and A. Volkov, Phys. Rev. B \textbf{58}, 15088 (1998).
\end{thebibliography}
\end{document}